\documentclass[pra,aps,epsfig,superscriptaddress,showpacs,twocolumn]{revtex4-1}
\usepackage{amsmath}
\usepackage{amssymb}
\usepackage{graphicx}
\usepackage{breakurl}
\usepackage{bm}
\usepackage{amsfonts}
\usepackage{mathrsfs}
\usepackage[unicode=true,pdfusetitle,pdfborder={0 0 1}]{hyperref}

\hypersetup{colorlinks,
linkcolor=blue,          citecolor=blue,        filecolor=blue,      urlcolor=blue           }
\graphicspath{{../figures/}}

\begin{document}

\title{Quantized electromagnetic response of three-dimensional chiral
topological insulators}

\author{S.-T. Wang}
\affiliation{Department of Physics, University of Michigan, Ann Arbor, Michigan 48109, USA}
\affiliation{Center for Quantum Information, IIIS, Tsinghua University, Beijing 100084,
PeopleÕs Republic of China}

\author{D.-L. Deng}
\affiliation{Department of Physics, University of Michigan, Ann Arbor, Michigan 48109, USA}
\affiliation{Center for Quantum Information, IIIS, Tsinghua University, Beijing 100084,
PeopleÕs Republic of China}

\author{Joel E. Moore}
\affiliation{Department of Physics, University of California, Berkeley, California 94720, USA}
\affiliation{Materials Sciences Division, Lawrence Berkeley National Laboratory, Berkeley, California 94720, USA}

\author{Kai Sun}
\affiliation{Department of Physics, University of Michigan, Ann Arbor, Michigan 48109, USA}

\author{L.-M. Duan}
\affiliation{Department of Physics, University of Michigan, Ann Arbor, Michigan 48109, USA}
\affiliation{Center for Quantum Information, IIIS, Tsinghua University, Beijing 100084,
PeopleÕs Republic of China}

\date{\today }

\begin{abstract}
Protected by the chiral symmetry, three dimensional chiral topological
insulators are characterized by an integer-valued topological invariant. How this invariant could emerge in physical observables is an important
question. Here we show that the magneto-electric polarization can identify the integer-valued invariant if we gap the system without coating a quantum Hall layer on the surface. The quantized response is demonstrated to be robust against weak perturbations. We also study the topological properties by adiabatically coupling two nontrivial phases, and find that gapless states appear and are localized at the boundary region. Finally, an experimental scheme is proposed to realize the Hamiltonian and measure the quantized response with ultracold atoms in optical lattices.    
\end{abstract}

\maketitle

\section{Introduction}

The periodic table of topological insulators (TIs) and superconductors
classifies topological phases of free fermions according to the system
symmetry and spatial dimensions \cite{Schnyder:2008ez, kitaev2009periodic}.
Notable examples include integer quantum Hall insulators breaking all those
classification symmetries and the time-reversal-invariant TIs protected by
the time-reversal symmetry \cite{hasan2010colloquium,moore2010birth,
Qi:2011wt,kane2005z,bernevig2006quantum, fu2007TI3D}. Mathematically,
these exotic states can be characterized by various topological invariants.
An interesting question is how to relate these invariants to physical
observables. For integer quantum Hall insulators, the Chern number ($\mathbb{%
Z}$ invariant) corresponds to the quantized Hall conductance \cite%
{thouless1982quantized}, while for the time-reversal-invariant TIs, the $%
\mathbb{Z}_{2}$ invariant is associated with a quantized magneto-electric
effect in three dimensions (3D) \cite{Qi:2008eu, Essin:2009ui}. 

The 3D chiral TIs protected by the chiral symmetry \cite%
{Hosur:2010ie, Ryu:2010ko} are of particular interest as they are 3D TIs
characterized by a $\mathbb{Z}$ (instead of $\mathbb{Z}_{2}$) invariant and
may be realized in ultracold atomic gases with engineered spin-orbital
coupling \cite{Lewenstein:2007hr, dalibard:2011gg, goldman2013light}. An experimental scheme was recently proposed to implement a three-band chiral TI in an optical lattice \cite{Wang2014probe}. For such 3D chiral TIs, it is known that the topological magneto-electric effect should also arise, but in theory it captures only the $\mathbb{Z}_{2}$ part of the $\mathbb{Z}$ invariant due to the gauge dependence of the polarization in
translationally invariant systems \cite{Hosur:2010ie}. It is thus an
important question to find out how the $\mathbb{Z}$ character could manifest
itself in experiments. It was proposed in Ref.\ \cite%
{shiozaki2013electromagnetic} that the $\mathbb{Z}$ effect may become
visible in certain carefully engineered heterostructures, but the
implementation of such a heterostructure is experimentally challenging.

In this paper, we study the nontrivial $\mathbb{Z}$ character of the chiral
TI by exploring the adiabatic transition between two nontrivial phases and
by numerically simulating the magneto-electric effect in a single phase. We show that not only
the $\mathbb{Z}_{2}$ response but the $\mathbb{Z}$ character can be observed
by gapping the system without adding a quantum Hall layer on the surface, i.e., the
ambiguity resulting from different terminations appears to be avoidable in practice.
Also, the quantized polarization is demonstrated to be robust against small perturbations even in the absence of a perfect chiral symmetry. This observation is important for experimental realization, because in a real system the chiral symmetry is typically an approximate instead of exact symmetry. Lastly, we propose an experimental scheme to realize the Hamiltonian and probe the integrally quantized response with cold atomic systems.

\section{Model and Topological Characterization} \label{Sec:Model}

We first introduce a minimal lattice tight-binding model for chiral
topological insulators with the Hamiltonian in the momentum space given by $%
H_{1}=\sum_{\mathbf{k}}\Psi _{\mathbf{k}}^{\dagger }\mathcal{H}_{1}(\mathbf{k%
})\Psi _{\mathbf{k}}$, where $\Psi _{\mathbf{k}}=(a_{\mathbf{k\uparrow }},a_{%
\mathbf{k}\downarrow },b_{\mathbf{k}\uparrow },b_{\mathbf{k}\downarrow })^{T}
$ denotes fermionic annihilation operators with spins $\uparrow ,\downarrow $
on sublattices or orbitals $a,b$. In cold atom systems, the pseudospins and
orbitals can be represented by different atomic internal states. The $%
4\times 4$ Hamiltonian is 
\begin{equation}
\mathcal{H}_{1}(\mathbf{k})\!=\! \left( \! 
\begin{array}{cccc}
0 & 0 & -iq_{0}+q_{3} & q_{1}-iq_{2} \\ 
0 & 0 & q_{1}+iq_{2} & -iq_{0}-q_{3} \\ 
iq_{0}+q_{3} & q_{1}-iq_{2} & 0 & 0 \\ 
q_{1}+iq_{2} & iq_{0}-q_{3} & 0 & 0%
\end{array}%
\! \! \right) 
\end{equation}%
where $q_{0}=h+\cos k_{x}+\cos k_{y}+\cos k_{z},q_{1}=\sin k_{x}+\delta
,q_{2}=\sin k_{y},q_{3}=\sin k_{z}$, with $h,\delta $ being control
parameters. The lattice constant and tunneling energy are set to unity. In
real space, this Hamiltonian represents on-site and nearest neighbor
hoppings and spin-flip hoppings between two orbitals in a simple cubic lattice. These hoppings can be
realized by two-photon Raman transitions in cold atoms \cite%
{jaksch2003creation,Aidelsburger:2013ew,Miyake:2013jw,Wang2014probe}. The
energy spectrum for this Hamiltonian is $E_{\pm }(\mathbf{k})=\pm \lbrack
(\sin k_{x}+\delta )^{2}+\sin ^{2}k_{y}+\sin ^{2}k_{z}+(\cos k_{x}+\cos
k_{y}+\cos k_{z}+h)^{2}]^{1/2}$, with two-fold degeneracy at each $\mathbf{k}
$. For $\delta =0$, the system acquires time-reversal symmetry $T$,
particle-hole symmetry $C$, and chiral symmetry $S=TC$, which can be explicitly seen as~\cite{Ryu:2010ko}:
\begin{eqnarray}
T:&  \quad  (\sigma _{x}\otimes \sigma _{y})[\mathcal{H}_{1}(\mathbf{k}%
)]^{\ast }(\sigma _{x}\otimes \sigma _{y}) &=\mathcal{H}_{1}(-%
\mathbf{k}) \\
C:&  \quad   (\sigma _{y}\otimes \sigma _{y})[\mathcal{H}_{1}(\mathbf{k}%
)]^{\ast }(\sigma _{y}\otimes \sigma _{y}) &=-\mathcal{H}_{1}(-\mathbf{k}) \\
S:&  \quad  (\sigma _{z} \otimes I_{2})[\mathcal{H}_{1}(\mathbf{k}%
)](\sigma _{z} \otimes I_{2}) &=-\mathcal{H}_{1}(\mathbf{k}) \label{eqn:ChiralSym}
\end{eqnarray}
with $\sigma_{i}$ as Pauli matrices, and $ I_{2}$ as the $2 \times 2$ identity matrix. When $\delta \neq 0$, time-reversal and particle-hole symmetries are
explicitly broken, but the chiral symmetry survives.

\begin{figure}[t]
\includegraphics[width=0.235\textwidth]{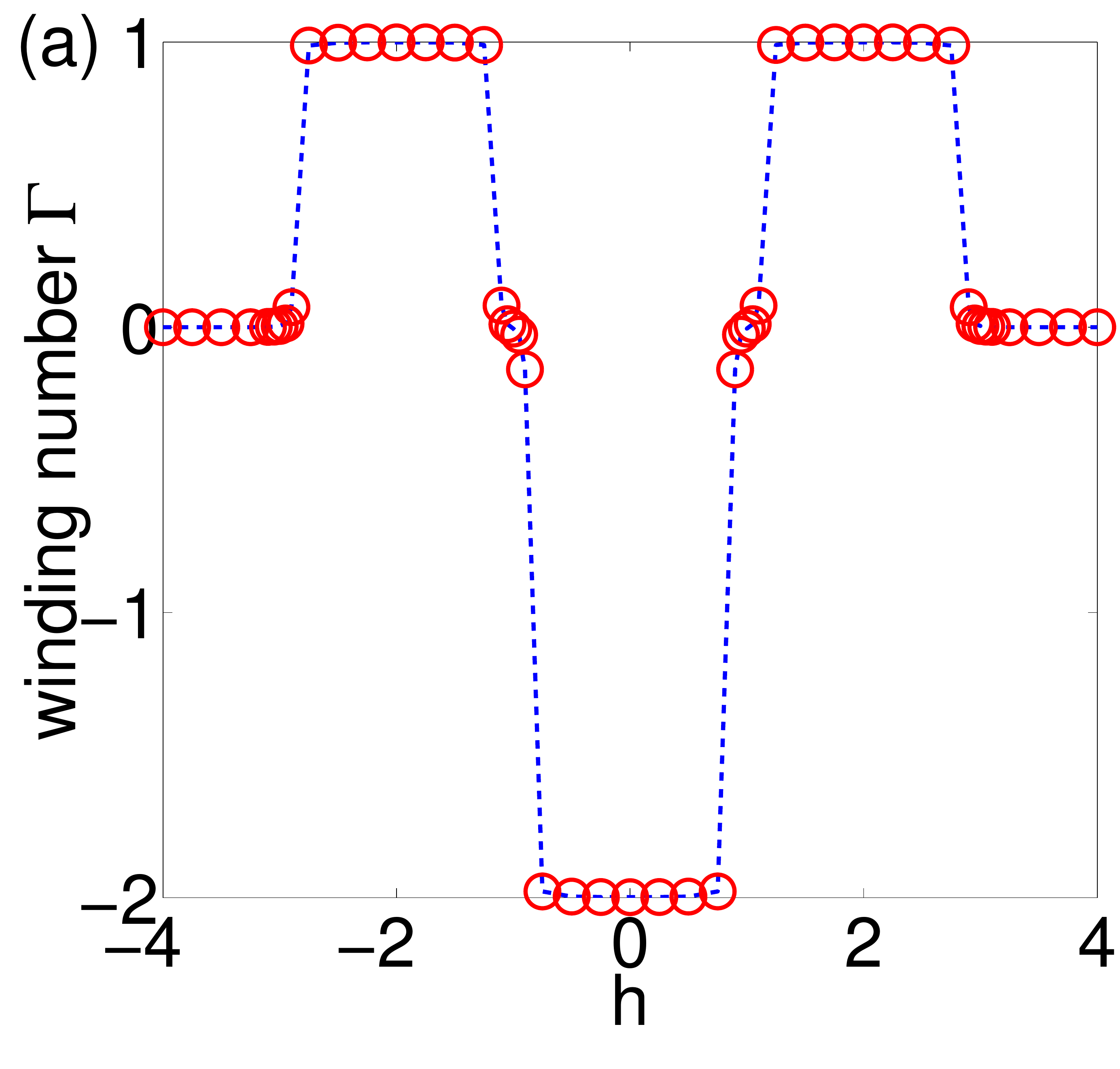} %
\includegraphics[width=0.235\textwidth]{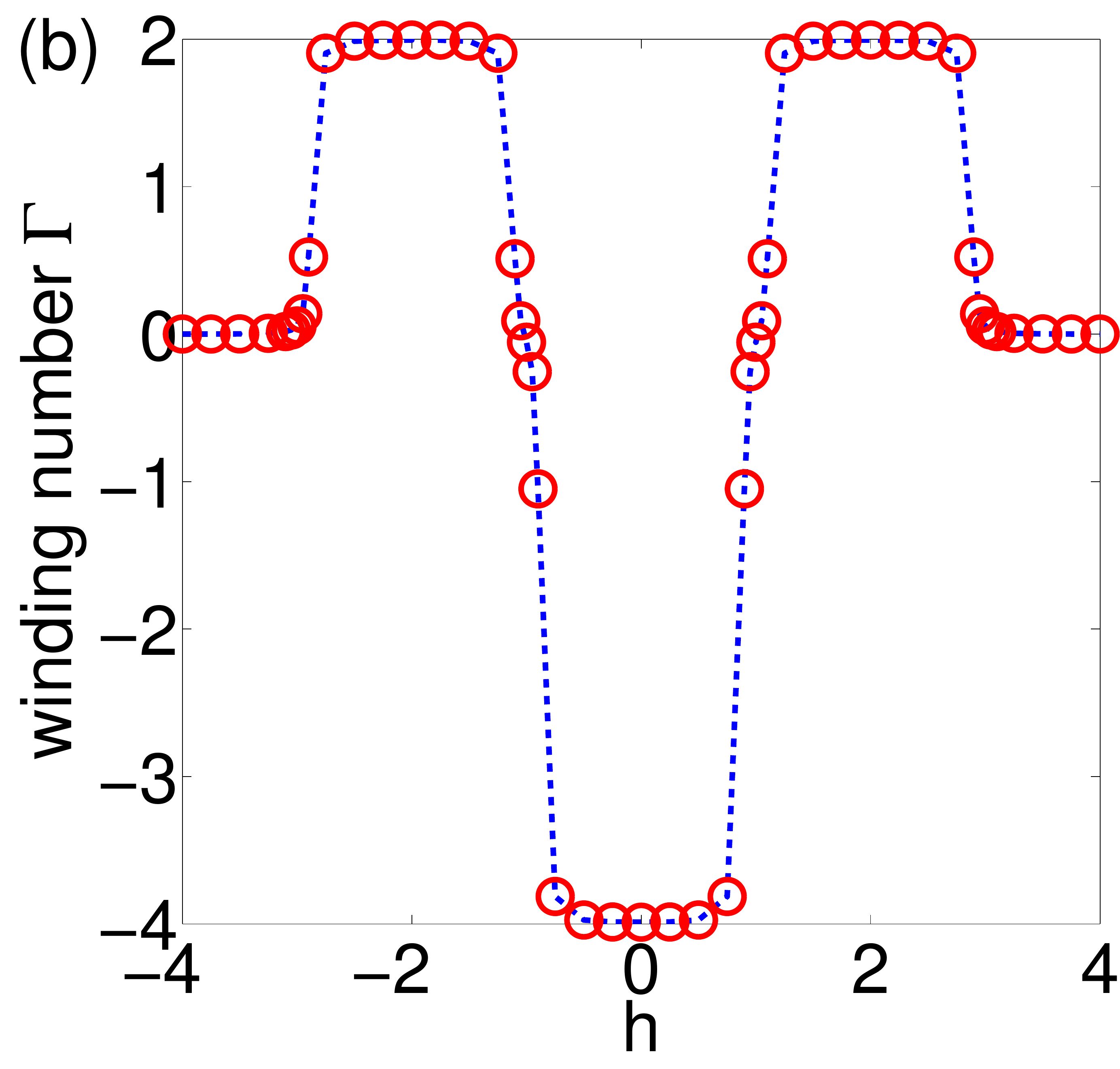}
\caption{(Color online) The winding number $\Gamma$ as a function of the parameter $h$. The
Hamiltonians are $\mathcal{H}_{1}(k)$ in (a) and $\mathcal{H}_{2}(k)$ in (b)
respectively. $\protect\delta=0.5$ for both panels.}
\label{fig:Index}
\end{figure}

The topological property of the Hamiltonian $\mathcal{H}_{1}(\mathbf{k})$ can be characterized by the winding number $\Gamma$  \cite{Schnyder:2008ez, Ryu:2010ko}. To define the winding number, let us introduce the $Q$ matrix, $Q(\mathbf{k})=1-2P(\mathbf{k})$, where $P(\mathbf{k})=\sum_{f}|u_{f}(\mathbf{k}
)\rangle \langle u_{f}(\mathbf{k})|$ is the projector onto the filled Bloch bands with wave-vectors $|u_{f}(\mathbf{k})\rangle$. The $Q$ matrix can be brought into the block off-diagonal form $Q( \mathbf{k)}=
\left(
\begin{array}{cc}
0 & b(\mathbf{k}) \\
b^{\dagger }(\mathbf{k}) & 0%
\end{array}%
\right) $ with the chiral symmetry.
With the matrix $b(\mathbf{k})$, one can write \cite{Schnyder:2008ez,Ryu:2010ko}
\begin{equation}
\Gamma \! = \! \frac{1}{24\pi ^{2}}\int_{\text{BZ}}d\mathbf{k}%
\;\epsilon ^{\mu \rho \lambda }\text{Tr}[(b^{-1}\partial _{\mu
}b)(b^{-1}\partial _{\rho }b)(b^{-1}\partial _{\lambda }b)],
\label{Chiral-invariant}
\end{equation}%
where $\epsilon ^{\mu \rho \lambda }$ is the antisymmetric Levi-Civita
symbol and $\partial _{\mu }b\equiv \partial _{k_{\mu }}b(\mathbf{k})$. The Hamiltonian $\mathcal{H}_{1}(\mathbf{k})$ supports topological phases with $\Gamma =1,-2$. To obtain chiral TIs with arbitrary integer topological invariants, one can use the quaternion construction proposed in Ref.\ \cite{deng2013systematic}. By considering $q=q_{0}+q_{1}\boldsymbol{i}+q_{2}\boldsymbol{j}+q_{3}\boldsymbol{k}$ as a quaternion and raising to a power, all $\mathbb{Z}$ topological phases can be realized by the family of tight-binding Hamiltonians. By taking the quaternion square, for example, one
obtains $q^{2}=q_{0}^{2}-q_{1}^{2}-q_{2}^{2}-q_{3}^{2}+2q_{0}q_{1}%
\boldsymbol{i}+2q_{0}q_{2}\boldsymbol{j}+2q_{0}q_{3}\boldsymbol{k}$, and we
can therefore acquire another tight-binding Hamiltonian $H_{2}=\sum_{\mathbf{%
k}}\Psi _{\mathbf{k}}^{\dagger }\mathcal{H}_{2}(\mathbf{k})\Psi _{\mathbf{k}}
$ with each component of $q^{2}$ replacing the respective components $%
q_{0},q_{1},q_{2},q_{3}$ in $\mathcal{H}_{1}(\mathbf{k})$. This second Hamiltonian $%
\mathcal{H}_{2}(\mathbf{k})$ contains next-nearest-neighbor hopping terms.
The winding number $\Gamma$ can be calculated numerically by discretizing the Brillouin zone and replacing the integral by a discrete sum \cite{deng2013systematic}. The results are shown in Fig.\ \ref{fig:Index} for both $\mathcal{H}_{1}(\mathbf{k})$ and $\mathcal{H}_{2}(\mathbf{k})$.

\begin{figure}[b]
\includegraphics[width=0.5\textwidth]{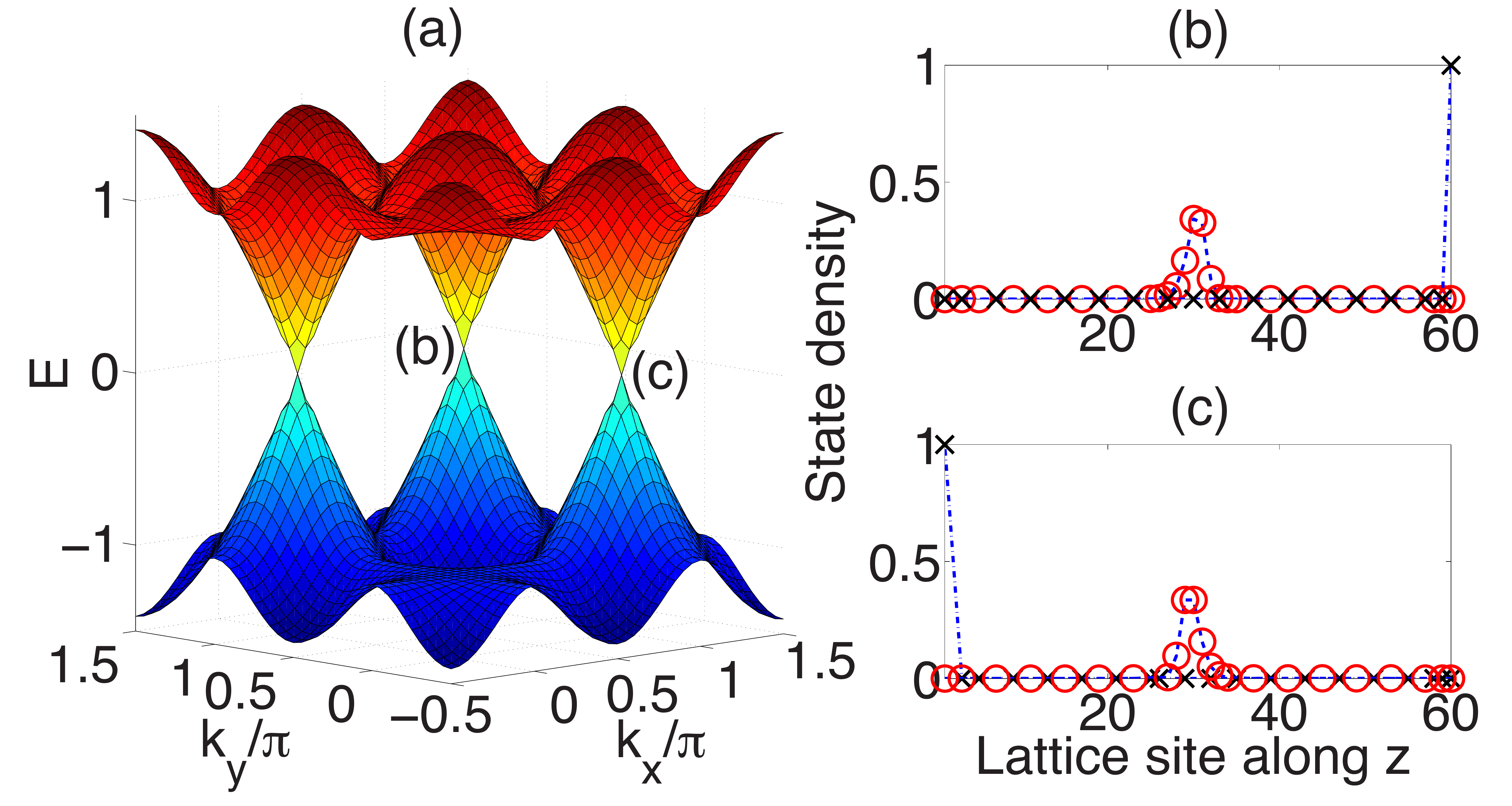}
\caption{(Color online) Coupling two topologically nontrivial phases by varying the parameter $h$ adiabatically from 0 to 2 along the $z$ direction. $x$ and $y$ directions are periodic. Sixty slabs are taken for the $z$ direction. (a) The
energy dispersion for the lowest conduction band and highest valence band.
(b) and (c) show the surface states near the respective Dirac points. The $%
\Gamma$ point is displaced from the center for a better display of the Dirac
cones.}
\label{fig:Phase}
\end{figure}

\section{Surface states and Heterostructure of nontrivial topological phases}

By imposing an open boundary condition along the $z$ direction, and keeping the $x$ and $y$ directions in momentum space, surface Dirac cones are formed for nontrivial topological phases. We find that the winding number coincides with the total number of Dirac cones counted for all inequivalent surface states (i.e.\ not counting the twofold degeneracy for each band), which
confirms explicitly the bulk-edge correspondence (See Appendix \ref{App:A}). A
distinctive difference from the time-reversal invariant TI is that any
number of Dirac cones on the surface is protected by the chiral symmetry 
\cite{Hosur:2010ie}.

With an integer number of nontrivial phases, it is intriguing to study the
topological phase transition between two different phases. A simple way to
explore this is to adiabatically vary $h$ from one end to the other end of
the sample. The parameter $h$ concerns the onsite tunneling strength between
opposite orbitals ($a_{\mathbf{\uparrow }}^{\dag }b_{\mathbf{\uparrow }}$
and $a_{\mathbf{\downarrow }}^{\dag }b_{\mathbf{\downarrow }}$ terms). This
hopping can be realized by a two-photon Raman process, and the strength $h$
can be controlled by the laser intensity \cite{Wang2014probe}. Numerically,
we vary $h$ in the form of $h=1+\tanh (z-L_{z}/2)$, where $z$ denotes the $z$th layer and $L_{z}$ the total number of slabs along the $z$ direction. This
form ensures that $h$ changes adiabatically from $0$ on one end to $2$ on
the other end of the sample, so that it effectively couples two nontrivial
phases. For the Hamiltonian $\mathcal{H}_{1}$, it couples two topological
phases with winding numbers $\Gamma =-2$ and $\Gamma =1$. Similar to the
interface between a topological insulator and the trivial vacuum, a surface
state should appear at the interface. As shown in Fig.\ \ref{fig:Phase},
three Dirac cones are formed inside the band gap. In addition to the surface
states observed on both ends of the sample, a localized state is formed at
the interface between two topologically distinct regions. These interface
states are always present regardless of the detailed structure of the
interface. Even for sharp boundaries, the interface states remain. The Dirac
cone structure may be probed through Bragg spectroscopy in cold atom
experiments \cite{stamper1999excitation,zhu2007simulation}.

The above heterostructure could be used to probe the topological properties of the chiral TI, but it is experimentally difficult to engineer such a heterostructure, especially in cold atom systems. In Ref.~\cite{shiozaki2013electromagnetic}, it was shown that the $\mathbb{Z}$ character of the topological invariant could be seen in some carefully engineered heterostructures. In the following, we show that the $\mathbb{Z}$ topological invariant can be observed via the magneto-electric polarization in a single nontrivial phase with a gapped surface. We further show that the detection will be robust to realistic experimental perturbations and present a feasible experimental scheme to observe the quantized response.   

\section{Magneto-electric effect}

\begin{figure*}[t]
\includegraphics[width=\textwidth]{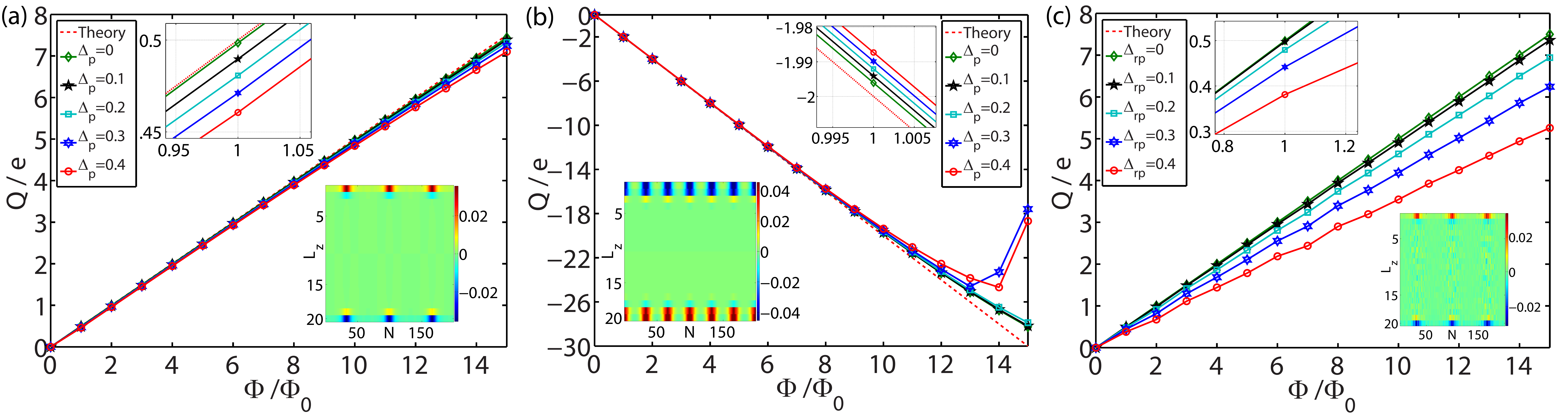}
\caption{(color online). Charge $Q$ accumulated on the surface in the $z$ direction due to a uniform magnetic field with total flux $\Phi$ at $k_{x}=k_{y}=0$. $%
N=199,L_{z}=20$ for all three panels. (a) and (b) consider the perturbative
effect of an intra-site nearest neighbor hopping. (c) adds a random onsite
potential characterized by $\Delta_{\text{rp}}$. The insets show the charge density at $\Phi=3\Phi_{0}$ without perturbations, and a closeup for the slope. The parameters in each panel are: (a) Hamiltonian $\mathcal{H}_{1}(\mathbf{k})$ with $h=2,\protect\delta%
=0.5, \Delta_{S}=1, \Gamma=1$; (b) Hamiltonian $\mathcal{H}_{2}(\mathbf{k})$
with $h=0,\protect\delta=0, \Delta_{S}=1, \Gamma=-4$; (c) Hamiltonian $%
\mathcal{H}_{1}(\mathbf{k})$ with $h=2,\protect\delta=0, \Delta_{S}=1,
\Gamma=1$. In all panels, the accumulated charge $Q$ is relative to the case
when $\Phi=0$ and is summed over all particle species for the upper half of
the sample at half fillings.}
\label{fig:ChargeAccum}
\end{figure*}

The magneto-electric effect is a remarkable manifestation of the bulk non-trivial topology. The linear response of a TI to an electromagnetic field can be described by the magneto-electric polarizability tensor as~\cite{Essin:2009ui}
\begin{equation}
\alpha _{ij}=\dfrac{\partial P_{i}}{\partial B_{j}}\bigg|_{E=0}=\dfrac{%
\partial M_{j}}{\partial E_{i}}\bigg|_{B=0},
\end{equation}%
where $E$ and $B$ are the electric and magnetic field, $P$ and $M$ are the
polarization and magnetization. Unique to topological
insulators is a diagonal contribution to the tensor with $\alpha
_{ij}=\theta \frac{e^{2}}{2\pi h}\delta _{ij}$. This is a peculiar
phenomenon as an electric polarization is induced when a magnetic field is
applied along the same direction \cite{Qi:2008eu}. This effect can be
described by a low-energy effective field theory in the Lagrangian as ($c=1$)
\begin{equation}
\Delta \mathcal{L}=\theta \dfrac{e^{2}}{2\pi h}\mathbf{B}\cdot \mathbf{E}
\end{equation}%
known as the \textquotedblleft axion electrodynamics\textquotedblright\ term 
\cite{Essin:2009ui}. For time-reversal-invariant TIs, an equivalent
understanding will be a surface Hall conductivity induced by the bulk
magneto-electric coupling. When the time-reversal symmetry is broken on the
surface generating an insulator, a quantized surface Hall conductance will
be produced: 
\begin{equation}
\sigma _{H}=\theta \frac{e^{2}}{2\pi h}
\end{equation}%
where $\theta $ is quantized to be $0$ or $\pi $ to preserve the
time-reversal invariance \cite{Qi:2008eu}. $\theta =\pi $ corresponds to the
non-trivial time-reversal-invariant TI with a fractional quantum Hall
conductivity. The electric polarization can be understood with Laughlin's
flux insertion argument \cite{Laughlin:1981jd}. A changing magnetic field
through the insulator induces an electric field (by Faraday's law), which
together with the quantized Hall conductivity will produce a transverse
current and accumulate charge around the magnetic flux tube at a rate
proportional to $\sigma _{H}$ as $Q=\sigma _{H} \Phi $ \cite%
{rosenberg2010wormhole}.

\subsection{Numerical results}

Theoretically, chiral TIs are also predicted to have this topological
magneto-electric effect \cite{Hosur:2010ie,Ryu:2010ko}. The field theory
only captures the $\mathbb{Z}_{2}$ part of the integer winding number due to
the $2\pi $ periodicity of $\theta $ associated with a gauge freedom in
transitionally invariant systems. However, we numerically show that the $%
\mathbb{Z}$ character can actually be observed by gapping the system without
adding a strong surface orbital magnetic field. Apparently, this corresponds
to a particular gauge such that the $\mathbb{Z}$ character can be distilled
from the polarization. More concretely, we consider the chiral TI
represented by both Hamiltonians $\mathcal{H}_{1}(\mathbf{k})$ and $\mathcal{%
H}_{2}(\mathbf{k})$. A uniform magnetic field is inserted through the chiral
TI sample via the Landau gauge $\mathbf{A}=Bx\hat{y}$ with a minimal
coupling by replacing $k_{y}$ with $k_{y}-\frac{e}{\hbar }Bx$. We keep $x$
and $y$ directions in momentum space, and the $z$ direction in real space
with open boundaries and $L_{z}$ slabs. By taking a magnetic unit cell with $%
N$ sites along the $x$ direction, the Hamiltonian can be partially Fourier
transformed to be a $4L_{z}\times N$ matrix for each $k_{x}$ and $k_{y}$,
with $4$ taking into account of spins $\uparrow ,\downarrow $ and orbitals $%
a,b$. For a unit magnetic cell with $N$ lattice cells, the total magnetic
flux through the unit cell is quantized to be integer multiples of a full
flux quantum $\Phi =n\Phi _{0}=n\frac{h}{e}$ due to the periodic boundary
condition along the $x$ direction, so the flux through a single lattice
plaquette is quantized to be $\Phi /N$. In the weak magnetic field limit,
one needs to take a large $N$. Besides the bulk Hamiltonian $\mathcal{H}_{1}(%
\mathbf{k})$ or $\mathcal{H}_{2}(\mathbf{k})$, we also add a surface term to
break the chiral symmetry and open a gap on the surface, 
\begin{equation}
H_{S} = \Delta _{S}  \sum_{k_{x},k_{y}} \sum_{j\in \text{surf}}\mathbf{\hat{S}}%
\cdot \hat{z}\left( \Psi _{j,k_{x},k_{y}}^{\dagger }\left( I_{2}%
\otimes \sigma _{z}\right) \Psi _{j,k_{x},k_{y}}\right),
\label{eqn:surfaceterm}
\end{equation}%
where $\mathbf{\hat{S}}$ represents the unit vector perpendicular to the
surface along $z$ direction, so $\mathbf{\hat{S}}\cdot \hat{z}=1$ for the
upper surface, and $\mathbf{\hat{S}}\cdot \hat{z}=-1$ for the lower surface.
This term represents a surface magnetization with a Zeeman coupling,
creating a different chemical potential for spins $\uparrow $ and $%
\downarrow $. It can be directly verified that this surface term breaks the
chiral symmetry $S$ in Eq.\ \eqref{eqn:ChiralSym}.

As the surface becomes gapped, at half filling, an increasing uniform
magnetic field accumulates charges on the surface via the magneto-electric
coupling. In the weak magnetic field limit, the charge accumulated on the
surface is proportional to $\sigma_{H}$ as 
\begin{equation}
Q= \sigma_{H} \Phi= \dfrac{\theta}{2\pi} n e.
\end{equation}
\emph{A priori}, $\theta$ needs not be quantized. However, analogous to the role played by time-reversal symmetry, chiral symmetry pins down $\theta$ to be $m\pi$ with an integer value $m$~\cite{Ryu:2010ko}. The fractional Hall conductivity, which cannot be removed by surface manipulations, emerges when $m$ is odd \cite{Qi:2008eu, Hosur:2010ie, Chang2013Experimental}. The integer part of $\sigma_{H}$, however, depends on the details of the surface \cite{Qi:2008eu,Essin:2009ui,Essin:2010gy}. An intuitive picture is that the $2 \pi$ ambiguity in $\theta$ results from the freedom to coat an integer quantum Hall layer on the surface, or equivalently to change the chemical potential and hence the Landau level occupancy of the surface in an orbital magnetic field. However, once a fixed surface Hamiltonian is defined, the adiabatic change in polarization associated with the increase in magnetic flux does have a physical meaning. This ambiguity can be avoided in cold atom systems, where the precise Hamiltonian can be engineered, allowing a direct link between the winding number $\Gamma$ and the charge accumulation rate $\theta/2\pi$.

Numerical results in Fig.\ \ref{fig:ChargeAccum} show that $\Gamma =\theta
/\pi $, which reveals that the magneto-electric polarization is a direct
indication of the non-trivial bulk topological phase characterized by the
integer winding number. To gain some intuition for why in our Hamiltonian
the value $\Gamma =\theta /\pi $ is observed, consider how a Zeeman term and
an orbital magnetic term produce different quantum Hall effects for a Dirac
fermion: the latter leads to Landau levels with many intervening gaps and a
chemical-potential dependence of the Hall effect, while the former leads to
a single gap and only one value of the Hall effect. The surface term (\ref%
{eqn:surfaceterm}) apparently acts more like a Zeeman field in leading to a
single gap and a unique value of the magneto-electric effect. We have
confirmed this intuition by adding a strong orbital magnetic field in a
single-layer Hamiltonian $\mathcal{H}_{1}(\mathbf{k})$ (see Appendix \ref{App:B}). Landau-level like bands are formed, and the charge accumulation rate is changed by an integer value by varying the chemical potential.

\subsection{Robustness to perturbations}

In real physical systems, the chiral symmetry may not be strictly observed.
We therefore consider the effect of weak perturbations to the charge
quantization. A natural term to add is an intra-site nearest neighbor
hopping term: 
\begin{equation}
\mathcal{H}_{p}(\mathbf{k})= \Delta_{p} (\cos k_{x} +\cos k_{y} + \cos k_{z} ) I_{4}.
\end{equation}
This term breaks the chiral symmetry and permits nearest-neighbor hoppings
within the same sublattice. Figures \ref{fig:ChargeAccum}(a) and \ref{fig:ChargeAccum}(b) show the
charge accumulation on the surface with increasing magnetic field for
various strengths of perturbations. The quantized slope is indeed robust to
small perturbations in the limit of weak magnetic field. Fig.\ \ref%
{fig:ChargeAccum}(c) takes into account of random onsite potential with
various perturbing strengths, and it again shows the robustness of the
topological effect. This includes the effect of a weak harmonic trap
typically present in cold atom systems. Note that strong perturbations
destroy the topological phase. We also performed similar calculations by
squeezing the entire uniform magnetic field into a single flux tube with open boundaries. It shows the same linear relationship between the surface charge accumulation and the magnetic flux. This indicates the uniformity of magnetic field is not crucial to observe the topological magneto-electric polarization, which may be an advantage to experimental realization.

\begin{figure*}[t]
\includegraphics[width=\textwidth]{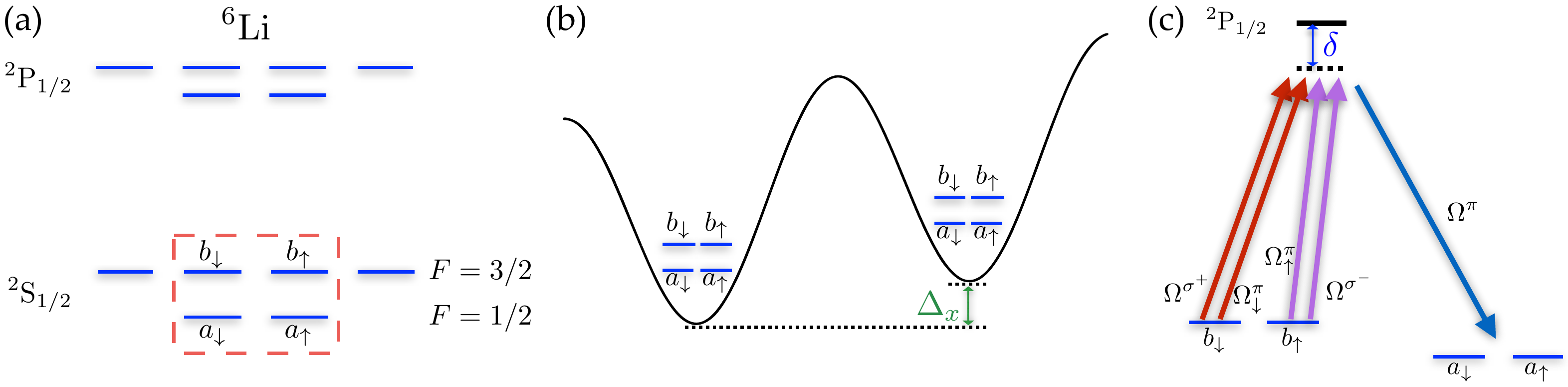}
\caption{(Color online) Schematics to realize the Hamiltonian $H_{1}$ with cold atoms. (a) Atomic level structure of $^{6}\text{Li}$ and the four internal states used to represent the spin and orbital degrees of freedom. (b) The optical lattice is tilted with a homogeneous energy gradient along each direction. (c) Laser configurations to create the first term in $H_{\mathbf{rx}}$. The superscript on each Rabi frequency denotes the polarization of the beam.}
\label{fig:Lattice}
\end{figure*}

\section{Experimental Implementation and Detection}

In this section, we present more details on the  implementation scheme with ultracold atoms. In Ref.\ \cite{Wang2014probe}, an experimental proposal for a three-band chiral TI was put forward. The realization scheme for the four-band
Hamiltonian studied here will be similar, with the atomic internal
states representing the spin and orbital degrees of freedom. In the previous sections, we studied two Hamiltonians $\mathcal{H}_{1}(\mathbf{k})$ and $\mathcal{H}_{2}(\mathbf{k})$. The latter involves next-nearest-neighbor hopping terms, which will be very challenging for experiment to engineer. In the following, we demonstrate, however, that the implementation of $\mathcal{H}_{1}(\mathbf{k})$ is possible with current technologies. $\mathcal{H}_{1}(\mathbf{k})$ supports topological phases with index $\Gamma = 0,1, -2$. It would be very exciting if experiment could simulate $\mathcal{H}_{1}(\mathbf{k})$ and probe its nontrivial topological properties via magneto-electric polarization. 

The Hamiltonian $\mathcal{H}_{1}(\mathbf{k})$ was written in momentum space in Sec.~\ref{Sec:Model}. The real space equivalent can be expressed as (for simplicity, we take $h=0$, $\delta=0$.)
\begin{align}   \label{eqn:RealSpaceH}
H_{1} &= \sum_{\mathbf{r}} H_{\mathbf{rx}} + H_{\mathbf{ry}} + H_{\mathbf{rz}}, \\
H_{\mathbf{rx}} &= -i/2 \left( a^{\dagger}_{\uparrow, \mathbf{r+x}} +
a^{\dagger}_{\downarrow, \mathbf{r+x}}
   \right)
   \left(
   b_{\uparrow, \mathbf{r}} + b_{\downarrow, \mathbf{r}} 
   \right) \nonumber \\
   & -i/2 \left( a^{\dagger}_{\uparrow, \mathbf{r-x}} -
a^{\dagger}_{\downarrow, \mathbf{r-x}}
   \right)
   \left(
   b_{\uparrow, \mathbf{r}} - b_{\downarrow, \mathbf{r}} 
   \right)
   + \text{H.c.}, \nonumber \\
   H_{\mathbf{ry}} &= -i/2 \left( a^{\dagger}_{\uparrow, \mathbf{r+y}} +
i a^{\dagger}_{\downarrow, \mathbf{r+y}}
   \right)
   \left(
   b_{\uparrow, \mathbf{r}} - i b_{\downarrow, \mathbf{r}} 
   \right) \nonumber \\
   & -i/2 \left( a^{\dagger}_{\uparrow, \mathbf{r-y}} - i
a^{\dagger}_{\downarrow, \mathbf{r-y}}
   \right)
   \left(
   b_{\uparrow, \mathbf{r}} + i b_{\downarrow, \mathbf{r}} 
   \right)
   + \text{H.c.}, \nonumber \\
   H_{\mathbf{rz}} &= -i a^{\dagger}_{\uparrow, \mathbf{r+z}} b_{\uparrow, \mathbf{r}} 
   -i a^{\dagger}_{\downarrow, \mathbf{r-z}} b_{\downarrow, \mathbf{r}} +\text{H.c.} \nonumber
\end{align}
where $\mathbf{x,y,z}$ represents a unit vector along the $ x,y,z $-direction of the cubic lattice, and $a_{\sigma,\mathbf{r}} $ $(b_{\sigma,\mathbf{r}})$  denotes the annihilation operator of the fermionic mode at the $a$ $(b)$ orbital and site $\mathbf{r}$ with the spin state $\sigma$. Basically, all terms in the Hamiltonian are some spin superpositions from one orbital hopping to another orbital. In the following, we take the fermionic species $^{6} \text{Li}$, for instance, to illustrate the implementation scheme. Other fermionic atoms can also be used with suitable atomic levels. We make use of four internal states of the hyperfine ground state manifold to carry two pseudospins and two orbitals as depicted in Fig.~\ref{fig:Lattice}(a). On top of the cubic optical lattice, a linear tilt is assumed along each direction to break the left-right symmetry as does the Hamiltonian [Fig.~\ref{fig:Lattice}(b)]. This linear tilt can be accomplished with the natural gravitational field, the magnetic field gradient, or the gradient of a dc- or ac-Stark shift \cite{jaksch2003creation, Aidelsburger:2013ew, Miyake:2013jw, Wang2014probe}. The hopping between orbitals can be activated by two-photon Raman transitions. Here, we show how to get the first term in the Hamiltonian $H_{\mathbf{rx}}$, which is  $-i/2 ( a^{\dagger}_{\uparrow, \mathbf{r+x}} + a^{\dagger}_{\downarrow, \mathbf{r+x}}  )
\left(   b_{\uparrow, \mathbf{r}} + b_{\downarrow, \mathbf{r}}  \right) $. Every other terms are of similar forms and can be likewise laser-induced. We may decompose it to four separate hoppings from $b$ states at site $\mathbf{r}$ to $a$ states at site $\mathbf{r+x}$. As shown in Fig.~\ref{fig:Lattice}(c), each of the hopping terms can be induced by a Raman pair: 
\begin{alignat*}{3}
( \Omega_{\uparrow}^{\pi}, \Omega^{\pi}) &\rightarrow -\tfrac{i}{2} \, a^{\dagger}_{\uparrow, \mathbf{r+x}} b_{\uparrow, \mathbf{r}}, \; %
(\Omega^{\sigma^{-}}, \Omega^{\pi}) &\rightarrow  -\tfrac{i}{2}  \, a^{\dagger}_{\downarrow, \mathbf{r+x}} b_{\uparrow, \mathbf{r}}  \\%
(\Omega^{\sigma^{+}}, \Omega^{\pi}) &\rightarrow  -\tfrac{i}{2}  \, a^{\dagger}_{\uparrow, \mathbf{r+x}} b_{\downarrow, \mathbf{r}}, \; %
(\Omega^{\pi}_{\downarrow}, \Omega^{\pi}) &\rightarrow  -\tfrac{i}{2}  \, a^{\dagger}_{\downarrow, \mathbf{r+x}} b_{\downarrow, \mathbf{r}} %
\end{alignat*}
The superscript on each Rabi frequency denotes the polarization of the respective beam. The relative phase and amplitude of the hoppings can be controlled by the laser beams. We have four free parameters here, $\Omega_{\uparrow}^{\pi}, \Omega^{\sigma^{-}}, \Omega^{\sigma^{+}}, \Omega^{\pi}_{\downarrow}$, each of which can be adjusted individually to yield the required configuration. These degrees of freedom ensure all other terms in the Hamiltonian can be produced in a similar way. One important aspect we need to be careful is that no spurious terms will be generated with undesired laser coupling. This is guaranteed by energy matching and polarization selection rules. In the undressed atomic basis, all four internal states are at different energies (split by a magnetic field for example), so the four beams coupling $b$ states to the excited states will not interfere with each other. In addition, the different detunings along each direction, $\Delta_{x,y,z}$, preempt the interference of beams that induce hoppings along different directions. There will be, however, some onsite spin-flipping terms induced by the laser beams, but those can be explicitly compensated by some r.f.\ fields.

The above scheme is hence able to engineer the Hamiltonian $H_{1}$ and is feasible with current technologies. The actual experiment will be challenging since many laser beams are involved with careful detunings. Nonetheless, all these beams can be drawn from the same laser with small frequency shifts produced by an acoustic or electric optical modulator. The uniform orbital magnetic field required to observe the topological polarization can be imprinted from the phase of the laser beams as an artificial gauge field \cite{jaksch2003creation, Aidelsburger:2013ew, Miyake:2013jw, Wang2014probe}. Lastly, the gap opening term in Eq.~\eqref{eqn:surfaceterm} can be created by extra laser beams focused on the surfaces, producing an effective Zeeman splitting. Other gap opening mechanisms on the surface should also work since the magneto-electric polarization is a bulk effect. The accumulated charge will be detectable from atomic density measurements \cite{nelson2007imaging, bakr2010probing, sherson2010single, Deng2014Direct}. Note that the density measurements do not need to be restricted to the surfaces, as a measurement for half of the sample produces good results, as shown in Fig.~\ref{fig:ChargeAccum}. One half of the sample can be removed to another state by shining a laser beam. The density on the other half of the sample can in turn be measured by time-of-flight imaging \cite{Deng2014Direct}.  As we have demonstrated in the previous section, the charge quantization is very robust to perturbations, so any weak perturbations introduced to the Hamiltonian, even those breaking the chiral symmetry in the bulk, should not alter detection results. 

\section{Conclusions}

In summary, we study the $\mathbb{Z}$ character of chiral topological
insulators by simulating the quantized magneto-electric effect of a nontrivial phase. We show that the $\mathbb{Z}$ character, not only the $\mathbb{Z}_{2}$ part, can be observed through magneto-electric polarization by properly gapping the system. An experimental scheme is also proposed for implementation and detection with cold atoms. This demonstrates explicitly how the topological invariant appears in physical observables for chiral TIs and will be important for experimental characterization.

\begin{acknowledgments}
S.T.W., D.L.D., and L.M.D. are supported by the NBRPC (973 Program) 2011CBA00300 (2011CBA00302), the IARPA MUSIQC program, the ARO and the AFOSR MURI program. J.E.M. acknowledges support from NSF DMR-1206515 and the Simons Foundation.  K.S. acknowledges support from NSF under Grant No.\ PHY1402971 and the MCubed program at University of Michigan.
\end{acknowledgments}

\appendix

\section{Bulk-edge correspondence} \label{App:A}

The bulk edge correspondence tells us that the bulk topological index should
have a surface manifestation, typically through the number of gapless Dirac
cones on the surface. This is generally verified for lower topological
index, such as 1 or 2. By imposing an open boundary condition along the $z$
direction for chiral topological insulators of different index, we find that
the winding number corresponds to the total number of Dirac cones counted
for all inequivalent surface states (i.e. not counting degeneracies). 

\begin{figure*}[t]
\begin{minipage}{0.4\textwidth}
(a)
\includegraphics[width=\textwidth]{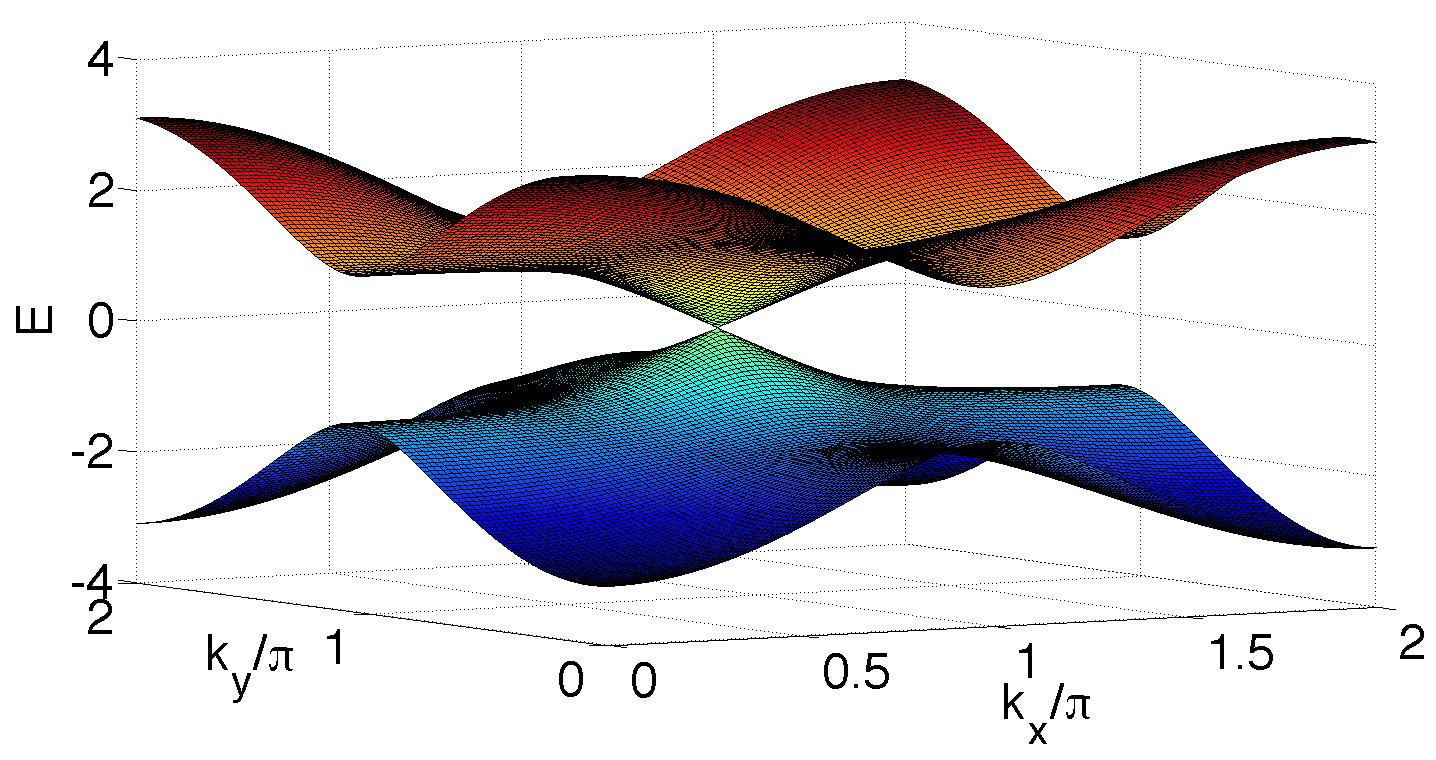} \\
\includegraphics[width=\textwidth]{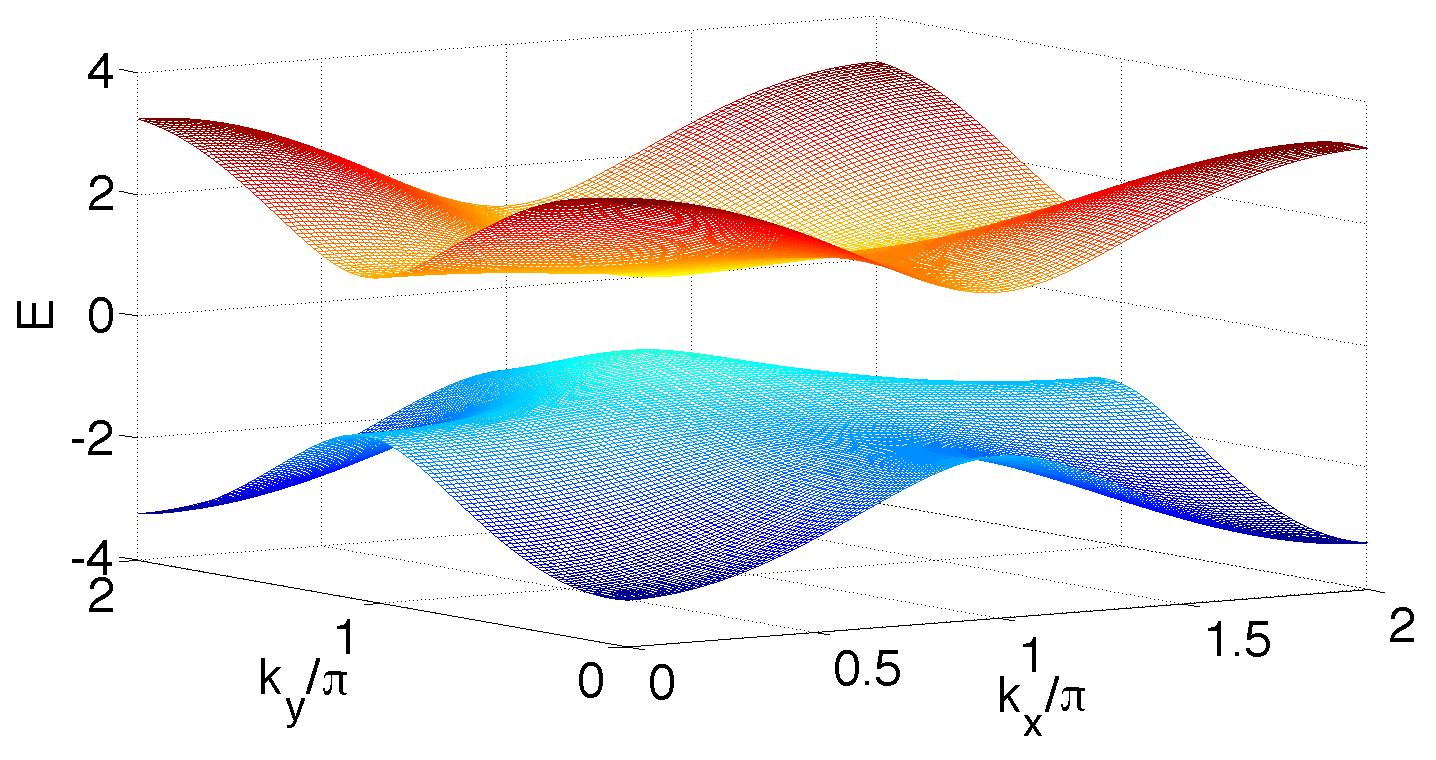}
\end{minipage}
\hspace{1.2cm} 
\begin{minipage}{0.4\textwidth}
(b)
\includegraphics[width=\textwidth]{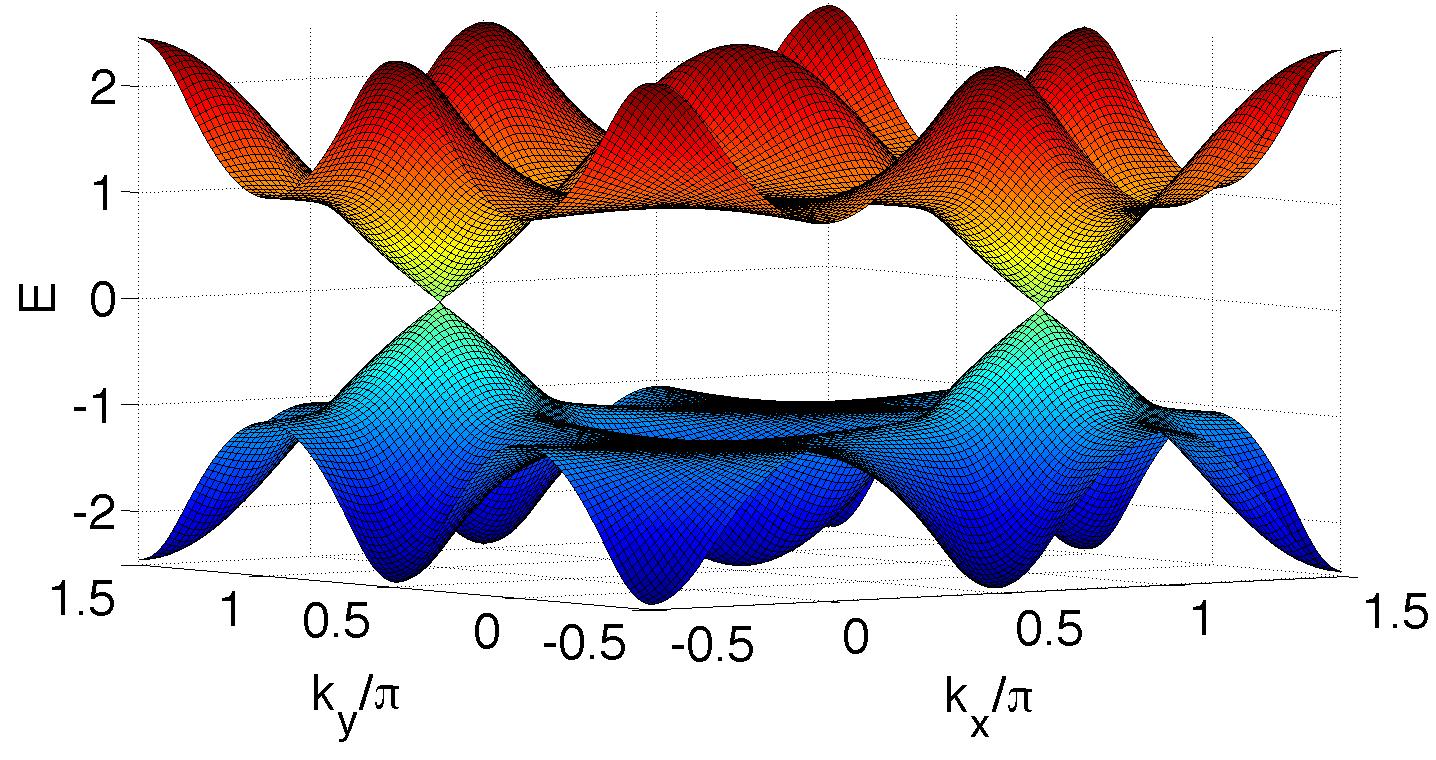} \\
\includegraphics[width=\textwidth]{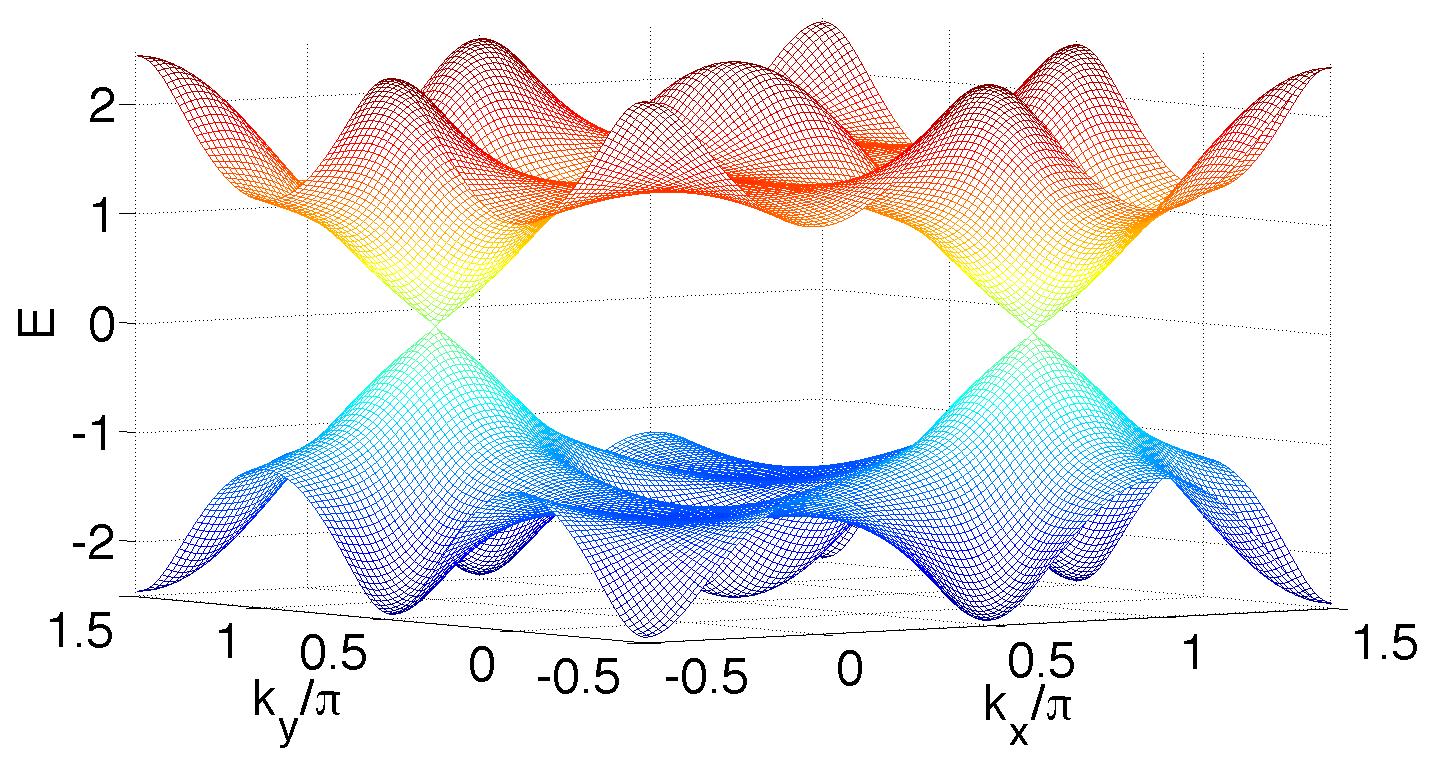}
\end{minipage}
\caption{(Color online) Spectrum for the surface states showing the number of Dirac cones.
The upper panels in (a) and (b) show the lowest conduction and highest
valence band. The lower panels show the next two bands closest to the Fermi
energy. (a) For $\mathcal{H}_{1}(\mathbf{k})$, $h=2, \protect\delta=0.5$
with winding number $\Gamma=1$ and 1 Dirac cone. (b) For $\mathcal{H}_{2}(%
\mathbf{k})$, $h=0,\protect\delta=0$ with winding number $\Gamma=4$ and 4
Dirac cones in total. The $\Gamma$ point is displaced from the center for
better display of the Dirac cones.}
\label{fig:Dirac}
\end{figure*}

Following Ref.\ \cite{deng2013systematic} to take a quaternion power $n$, we
can generalize the Hamiltonians in the main text from $\mathcal{H}_{1}(%
\mathbf{k})$ and $\mathcal{H}_{2}(\mathbf{k})$ to $\mathcal{H}_{n}(\mathbf{k}%
)$. For the Hamiltonian $\mathcal{H}_{1}(\mathbf{k})$ (i.e. $n=1$), when $h=2
$, the winding number $\Gamma=1$ guarantees the existence of 1 Dirac cone
[Fig.\ \ref{fig:Dirac}(a)]. For the Hamiltonian $\mathcal{H}_{2}(\mathbf{k})$
(i.e. $n=2$), when $h=0$, the winding number is $\Gamma=4$. So there are two
inequivalent surface states on each surface with two Dirac cones each [Fig.\ %
\ref{fig:Dirac}(b)]. In general, we have $m$ inequivalent surface
states with 1 Dirac cone each for  $n=m, 1<|h|<3, \Gamma=m$, and $m$ inequivalent surface states with 2 Dirac cones each for $n=m, -1<h<1, \Gamma=2m$. These have been explicitly verified up to $n=3$. Hence, the winding number $ \Gamma$ does correspond to the total number of Dirac cones for all inequivalent surface states.

\begin{figure*}[t]
\includegraphics[width=0.85\textwidth]{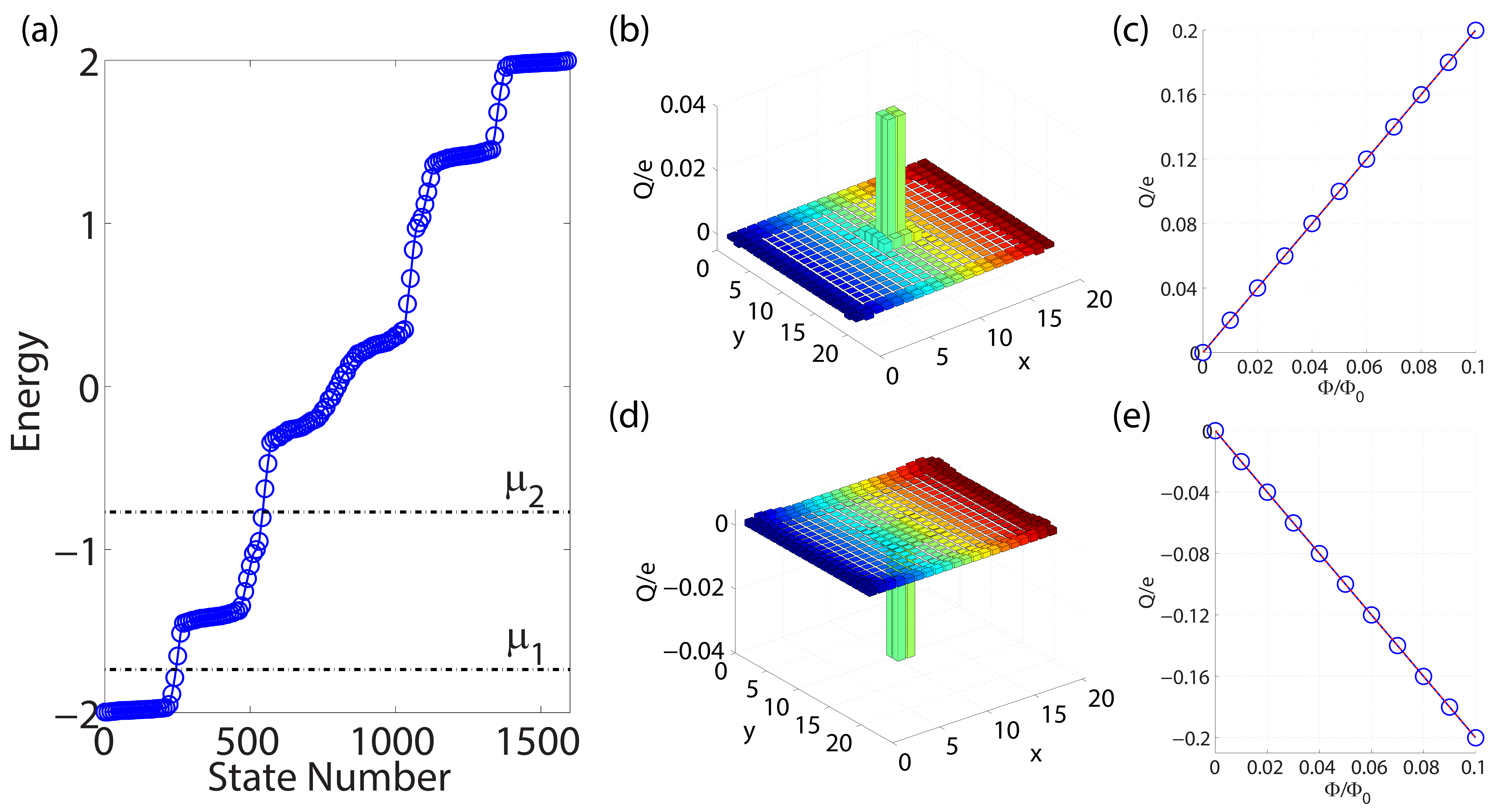}
\caption{(Color online) (a) Energy spectrum for the one-layer Hamiltonian $\mathcal{H}_{1}(%
\mathbf{k})$ with a strong uniform magnetic field and unit cell flux as $%
\frac{1}{3} \Phi_{0}$. An additional weak flux tube is inserted through the
center lattice cell of the layer, with flux up to $\Phi/\Phi_{0}=0.1$. (b)
and (c) [(d) and (e)] correspond to the charge polarization with respect to
the increasing flux tube at a chemical potential $\protect\mu_{1}$ [$\protect%
\mu_{2}$]. Charge is accumulated around the flux tube, and by changing the
chemical potential and hence the Landau level occupancy, the charge
accumulation rate can be modified by an integer.}
\label{Fig:Surf}
\end{figure*}

\section{Surface orbital field and integer quantum Hall layers} \label{App:B}

The $2\pi$ periodicity of the $\theta$ term is mathematically related to the
gauge freedom in the low-energy effective field theory. Physically, it is
associated with the freedom to coat an integer quantum Hall layer on the
surface, or equivalently to change the chemical potential and hence the
Landau level occupancy of the surface in an orbital magnetic field. Here, we
numerically verify this physical intuition. To do that, we consider a single
layer of the Hamiltonian $\mathcal{H}_{1}(\mathbf{k})$, so the $z$ component
drops out. A strong uniform orbital field is added to the layer via Peierls
substitution with the Landau gauge $\mathbf{A}= Bx \hat{y}$. $Ba^{2}= \frac{1%
}{3} \Phi_{0}$, where $a$ is the lattice constant, and $\Phi_{0}$ is the
flux quantum. For the Hofstadter Hamiltonian, this strong orbital field will
produce three gapped Landau levels. Here, a similar structure is developed
as shown in Fig.\ \ref{Fig:Surf}(a). There are six bands with the middle two
bands gapless. The extra number of bands are due to the spin and orbital
degrees of freedom. On top of the strong uniform magnetic field, an
additional weak flux tube is inserted through the center lattice. By
Laughlin's flux insertion argument, the charge accumulated around the flux
tube should be $Q/e = C \Phi/ \Phi_{0}$, where $C$ is the Chern number being
an integer. Figures \ref{Fig:Surf}(b) and \ref{Fig:Surf}(c) [\ref{Fig:Surf}(d) and \ref{Fig:Surf}(e)] show the charge polarization at
a chemical potential $\mu_{1}$ [$\mu_{2}$]. From the slope, we infer that
the first band has a Chern number $C=2$ and the second band has a Chern
number $C=-4$. So by changing the surface chemical potential, we could
modify the charge accumulation rate by an integer. Alternatively, in the
absence of this strong orbital magnetic field, with a surface gapping term $%
H_{S}$ in Eq.\ \eqref{eqn:surfaceterm} of the main text, there is only one gap and no such integer quantum Hall layers. Therefore the $\mathbb{Z}$ character of the winding number can be observed through such integrally quantized magneto-electric polarization measurements.


%

\end{document}